\ifdefined\kanjiskip
  \PassOptionsToPackage{dvipdfmx}{graphicx}
  \PassOptionsToPackage{dvipdfmx}{xcolor}
  \PassOptionsToPackage{dvipdfmx}{hyperref}
  \PassOptionsToPackage{backend=dvipdfmx}{expl3}
\fi
\documentclass[
    aps,
    prl,
    reprint,
    nobalancelastpage,
    floatfix,
    superscriptaddress,
    amsmath,
    amssymb,
    longbibliography
]{revtex4-2}

\usepackage{graphicx}
\usepackage{dcolumn}
\usepackage{bm}
\usepackage{mathtools}
\usepackage{physics}
\usepackage{braket}
\usepackage{amsthm}
\usepackage{xcolor}
\usepackage{booktabs}
\usepackage{hyperref}

\newcommand{\E}{\mathbb{E}}

\newcommand{\prlsubheading}[1]{%
  \par\indent\textnormal{#1:}\enspace\ignorespaces
}

\newcommand{\id}{\mathcal{I}}
\newcommand{\diamondnorm}[1]{\left\lVert #1 \right\rVert_{\diamond}}

\begin{document}

\title{Optimized Randomized Hamiltonian Simulation via Average-Error Analysis}

\author{Hayata Morisaki}
\email{u748119d@ecs.osaka-u.ac.jp}
\affiliation{
Graduate School of Engineering Science, The University of Osaka, 1-3 Machikaneyama, Toyonaka, Osaka 560-8531, Japan
}

\author{Keisuke Fujii}
\affiliation{
  Graduate School of Engineering Science, The University of Osaka, 1-3 Machikaneyama, Toyonaka, Osaka 560-8531, Japan
}
\affiliation{
  Graduate School of Informatics, Kyoto University, Sakyo-ku, Kyoto, 606-8501, Japan
}
\affiliation{
  Center for Quantum Information and Quantum Biology, The University of Osaka, 1-2 Machikaneyama, Toyonaka 560-0043, Japan
}
\affiliation{
  RIKEN Center for Quantum Computing (RQC), Hirosawa 2-1, Wako, Saitama 351-0198, Japan
}

%
%

\date{\today}

\begin{abstract}
Hamiltonian simulation is a central application of quantum computing.
Randomized Hamiltonian simulation approximates the target dynamics by sampling
quantum circuits and often allows simpler circuit implementations.
We develop a framework for randomized Hamiltonian simulation in which
Hamiltonian terms are sampled with arbitrary probabilities and the
corresponding short-time evolutions are implemented sequentially.
The leading channel error relative to ideal time evolution is governed by the
variance of the sampled generators.
Minimizing the variance-based upper bound on the worst-case error recovers
qDRIFT, a leading randomized Hamiltonian simulation algorithm that samples
terms in proportion to their operator norms.
For typical input states, however, this sampling distribution need not be
optimal.
Minimizing Haar-averaged error bounds instead yields sampling probabilities
proportional to the Hilbert--Schmidt norms of the terms.
This choice can provide smaller average-error bounds than conventional qDRIFT.
For Hamiltonians expressed as sums of Pauli strings, we combine this framework
with commuting Pauli grouping to obtain smaller error bounds than conventional
qDRIFT at fixed $R_z$ depth.
Numerical benchmarks for the Sachdev--Ye--Kitaev model show error-bound
improvement factors consistent with linear scaling in the number of qubits.
For the 108-qubit FeMoco Hamiltonian, the error bounds are reduced by a factor
of about 5.65.
These results establish average-error optimization as a practical design
principle for randomized Hamiltonian simulation.
\end{abstract}

\maketitle

\paragraph{Introduction.---}

Simulating the dynamics of quantum many-body systems is generally challenging
for classical computers because the Hilbert-space dimension grows exponentially
with system size.  Hamiltonian simulation---the implementation of
$U_t=e^{-itH}$ on a quantum computer---is a foundational primitive
of quantum computation~\cite{Feynman1982,Lloyd1996}.  Its applications include
studying quantum many-body dynamics and correlation functions, estimating
molecular energies~\cite{Lloyd1996,Reiher2017}, and preparing ground and thermal
states~\cite{Ge2019,PoulinWocjan2009}.  In particular, it is a key
subroutine of quantum phase estimation (QPE), which extracts spectral
information from time evolution~\cite{Kitaev1995,Cleve1998,AbramsLloyd1999}.

Hamiltonians are often written as a sum $H = \sum_{j=1}^K H_j$, where the time
evolution of each term is implementable.  A variety of methods have been
developed to construct the full evolution from such a decomposition, including
Trotter--Suzuki product formulas, linear-combination-of-unitaries (LCU) methods,
qubitization, and randomized simulation
algorithms~\cite{Lloyd1996,Berry2015,LowChuang2019,Campbell2019,LowChuang2017,LowKliuchnikovWiebe2019,ChildsOstranderSu2019,HaganWiebe2023,Nakaji2024}.
Among the randomized approaches, qDRIFT approximates the ideal time evolution
through a sequence of short-time evolutions generated by randomly sampled
terms.  Defining
$\lambda_{\mathrm{op}}:=\sum_{j=1}^{K}\norm{H_j}_{\mathrm{op}}$, standard qDRIFT
samples $H_j$ according to the operator-norm-proportional distribution
$p_j=\norm{H_j}_{\mathrm{op}}/\lambda_{\mathrm{op}}$ and applies a
correspondingly rescaled short-time evolution~\cite{Campbell2019}.  The required
number of qDRIFT steps depends on $\lambda_{\mathrm{op}}$ but not explicitly
on the number $K$ of terms.
This makes qDRIFT particularly attractive for quantum chemistry, where
Hamiltonians often contain many Pauli terms.

The operator-norm-proportional distribution is not, however, the only valid
choice for qDRIFT.  For any probabilities $p_j>0$, one may sample $H_j$ with
probability $p_j$ and evolve under $H_j/p_j$~\cite{Kiss2023}.  This implements importance sampling of $H$, since
$\sum_jp_j(H_j/p_j)=H$.
The average of these sampled short-time evolutions is therefore
a first-order approximation to the ideal short-time evolution.
The sampling distribution can be optimized to control higher-order
errors.  For example, Wu et al. proposed a state-dependent sampling
distribution designed to minimize the leading-order
infidelity~\cite{Wu2026}.

The standard qDRIFT distribution minimizes the usual
leading-order upper bound on the diamond distance to the ideal channel and
is thus tailored to a worst-case criterion~\cite{Campbell2019,Kiss2023}.
However, it does not necessarily minimize the error for typical input states.
Such worst-case bounds can be overly conservative in high-dimensional systems.
A similar gap between worst-case and typical errors has been observed for
product formulas: empirical Trotter errors can lie substantially below rigorous
worst-case bounds~\cite{SchubertMendl2023}, while average-case analysis with
random inputs replaces the operator norm by the normalized Hilbert--Schmidt
norm.  For a nearest-neighbor Heisenberg chain,
this changes the system-size scaling from $O(n)$ in the worst case to
$O(\sqrt n)$ on average~\cite{Zhao2022}.  Related concentration results give
improved typical-case gate counts for general local and number-conserving
fermionic Hamiltonians~\cite{ChenBrandao2024}.  For Trotter formulas,
average-error analysis can yield improved error bounds without changing the
simulation procedure.  For qDRIFT, however, changing the error criterion can also change
the preferred sampling distribution.  This raises two questions: how should
the sampling probabilities be chosen to control average errors, and how much
can the error bounds be improved?

In this work, we derive nonasymptotic upper bounds on the Haar-averaged
infidelity and absolute time-evolution signal error of qDRIFT with arbitrary
sampling probabilities.  Time-evolution signals contain information about
the energy spectrum of a Hamiltonian.  They can be measured using Hadamard
tests~\cite{Cleve1998} and used for single-ancilla phase
estimation~\cite{Kitaev1995,Dobsicek2007,WiebeGranade2016,OBrien2019,LinTong2022,Gunther2026}.
Both upper bounds are minimized by sampling in proportion to the normalized
Hilbert--Schmidt norms, $p_j^{(\mathrm{HS})}\propto\norm{H_j}_{\mathrm{HS}}$.
We use the normalized Hilbert--Schmidt norm
$\norm{A}_{\mathrm{HS}}:=\sqrt{\operatorname{Tr}(A^\dagger A)/d}$,
where $d$ is the Hilbert-space dimension.
With this sampling distribution, both average-error bounds are controlled by
$\lambda_{\mathrm{HS}}:=\sum_j\norm{H_j}_{\mathrm{HS}}$, rather than
$\lambda_{\mathrm{op}}$.  For a Hamiltonian expressed as a linear combination
of Pauli strings, $H=\sum_{\ell=1}^{L}c_\ell P_\ell$, sampling the terms individually
gives $\lambda_{\mathrm{op}}=\lambda_{\mathrm{HS}}=\sum_{\ell=1}^{L}\abs{c_\ell}$.
Thus both the worst-case and average-error bounds are controlled by the
coefficient $\ell_1$ norm.  When Pauli terms are grouped into blocks, each
treated as a single term in the Hamiltonian decomposition, the average-error
bounds instead depend on a sum of within-block
coefficient-$\ell_2$ norms.  Using a commuting Pauli-grouping method, we
numerically demonstrate error improvements over conventional qDRIFT at the
same $R_z$ depth for Sachdev--Ye--Kitaev (SYK)~\cite{SachdevYe1993,MaldacenaStanford2016}
and molecular Hamiltonians.
For the SYK model, the improvement factor in the error bounds grows
approximately linearly with the number of qubits.  For the 108-qubit FeMoco
Hamiltonian, we observe an approximately 5.65-fold improvement in the bounds.
Furthermore, for small systems, we confirm that the improvement factors in
the error bounds closely match those in the actual errors. 
These results establish average-error analysis as a design
principle for randomized Hamiltonian simulation.  The framework enables practical
classical evaluation of error bounds even for the 108-qubit FeMoco Hamiltonian,
and our benchmarks demonstrate reductions in these bounds at the same $R_z$
depth as conventional qDRIFT.

\paragraph{Importance-sampled qDRIFT.---}

We formulate importance-sampled qDRIFT~\cite{Kiss2023} with an arbitrary
sampling distribution and derive its leading channel error.  We then recover
the standard operator-norm sampling distribution from a worst-case
diamond-distance bound.
Consider an $n$-qubit Hamiltonian decomposed into Hermitian terms as
\begin{equation}
    H=\sum_{j=1}^{K}H_j,
    \qquad H_j=H_j^\dagger.
    \label{eq:hamiltonian-decomposition}
\end{equation}
We consider an arbitrary probability distribution satisfying
\begin{equation}
    p_j>0,
    \qquad \sum_{j=1}^{K}p_j=1.
    \label{eq:sampling-distribution}
\end{equation}
Divide the total evolution time $t$ into $N$ steps of duration
$\delta=t/N$.  For each step, choose $j$ with probability $p_j$ and apply
the evolution generated by $H_j/p_j$.  We define the corresponding
one-step qDRIFT channel by
\begin{equation}
    \mathcal{E}_{p,\delta}(\rho)
    :=\sum_{j=1}^{K}p_j
      e^{-i\delta H_j/p_j}\rho
      e^{i\delta H_j/p_j}.
    \label{eq:one-step-qdrift-channel}
\end{equation}
The sampled generator $H_j/p_j$ is an unbiased estimator of $H$, since
$\E_{j\sim p}[H_j/p_j]=H$.  We compare
Eq.~\eqref{eq:one-step-qdrift-channel} with the ideal short-time evolution
channel
\begin{equation}
    \mathcal{U}_{\delta}(\rho)
    :=e^{-i\delta H}\rho e^{i\delta H}.
    \label{eq:ideal-one-step-channel}
\end{equation}
To expand these channels, define the Liouvillians
\begin{equation}
    \mathcal{L}_j(\rho)=-i[H_j,\rho],
    \qquad
    \mathcal{L}=\sum_{j=1}^{K}\mathcal{L}_j,
    \label{eq:liouvillians}
\end{equation}
so that $\mathcal{U}_{\delta}=\exp(\delta\mathcal{L})$.
Expanding the qDRIFT and ideal channels gives
\begin{align}
    \mathcal{E}_{p,\delta}
    &=\id+\delta\mathcal{L}
      +\frac{\delta^2}{2}\sum_{j=1}^{K}
        \frac{\mathcal{L}_j^2}{p_j}
      +O(\delta^3),
      \\
    \mathcal{U}_{\delta}
    &=\id+\delta\mathcal{L}
      +\frac{\delta^2}{2}\mathcal{L}^2
      +O(\delta^3).
    \label{eq:one-step-expansions}
\end{align}
Thus importance reweighting makes the zeroth- and first-order terms agree
for every choice of positive sampling probabilities.  The leading local discrepancy is
governed by
\begin{equation}
    \mathcal{D}_p
    :=\sum_{j=1}^{K}\frac{\mathcal{L}_j^2}{p_j}
      -\mathcal{L}^2.
    \label{eq:qdrift-second-order-generator}
\end{equation}
For the total evolution time, we compose the one-step channel $N$ times:
\begin{equation}
    \mathcal{E}_{p,t}^{(N)}
    :=\mathcal{E}_{p,\delta}^{\,N}.
    \label{eq:average-qdrift-channel}
\end{equation}
For fixed $t$, $H$, and $p$, a telescoping expansion sums the
propagated one-step errors.  Writing this sum as a Riemann integral gives
\begin{equation}
    \mathcal{E}_{p,t}^{(N)}-\mathcal{U}_t
    =\frac{t^2}{2N}\int_0^1
      e^{(1-s)t\mathcal{L}}\mathcal{D}_p
      e^{st\mathcal{L}}\,\mathrm{d}s
      +O(N^{-2}).
    \label{eq:global-channel-difference}
\end{equation}
The conventional qDRIFT distribution follows from controlling the
worst-case channel error.  Since the diamond norm is invariant under
composition with unitary channels and
$\diamondnorm{\mathcal{L}_j}\leq2\norm{H_j}_{\mathrm{op}}$, Eq.~\eqref{eq:global-channel-difference}
implies
\begin{align}
    \diamondnorm{\mathcal{E}_{p,t}^{(N)}-\mathcal{U}_t}
    &\leq\frac{t^2}{2N}\diamondnorm{\mathcal{D}_p}
      +O(N^{-2})
      \nonumber\\
    &\leq\frac{2t^2}{N}
      \left[
        \sum_{j=1}^{K}\frac{\norm{H_j}_{\mathrm{op}}^2}{p_j}
        +\norm{H}_{\mathrm{op}}^2
      \right]
      +O(N^{-2}).
    \label{eq:diamond-distance-bound}
\end{align}
The Cauchy--Schwarz inequality gives
\begin{align}
    \sum_{j=1}^{K}\frac{\norm{H_j}_{\mathrm{op}}^2}{p_j}
    &=\left(
      \sum_{j=1}^{K}\frac{\norm{H_j}_{\mathrm{op}}^2}{p_j}
      \right)
      \left(\sum_{j=1}^{K}p_j\right)
      \nonumber\\
    &\geq\left(
      \sum_{j=1}^{K}
      \frac{\norm{H_j}_{\mathrm{op}}}{\sqrt{p_j}}\sqrt{p_j}
      \right)^2
      \nonumber\\
    &=\left(\sum_{j=1}^{K}\norm{H_j}_{\mathrm{op}}\right)^2.
    \label{eq:diamond-objective}
\end{align}
Thus, the upper bound in Eq.~\eqref{eq:diamond-distance-bound} is minimized by
the conventional qDRIFT distribution~\cite{Campbell2019},
\begin{equation}
    p_j^{(\mathrm{op})}
    =\frac{\norm{H_j}_{\mathrm{op}}}{\lambda_{\mathrm{op}}},
    \qquad
    \lambda_{\mathrm{op}}:=\sum_{k=1}^{K}\norm{H_k}_{\mathrm{op}}.
    \label{eq:standard-qdrift-distribution}
\end{equation}
For this distribution, using
$\norm{H}_{\mathrm{op}}\leq\lambda_{\mathrm{op}}$ gives
\begin{equation}
    \diamondnorm{\mathcal{E}_{p^{(\mathrm{op})},t}^{(N)}-\mathcal{U}_t}
    \leq\frac{4\lambda_{\mathrm{op}}^2t^2}{N}+O(N^{-2}).
    \label{eq:standard-qdrift-error-bound}
\end{equation}
This worst-case criterion need not be optimal for the average or
task-specific errors relevant to a particular experiment.  We therefore
retain the freedom to choose $p$ below.

\paragraph{Average-error-optimized sampling.---}
We now turn from the worst-case diamond distance to two Haar-averaged error
measures: the infidelity and the time-evolution signal error.
For both quantities, we derive nonasymptotic bounds governed by the common variance parameter $\operatorname{Tr}(D_p)/d$, defined below.  Minimizing this parameter yields the
normalized Hilbert--Schmidt-norm sampling rule
$p_j^{(\mathrm{HS})}\propto\norm{H_j}_{\mathrm{HS}}$, which can differ from the
conventional operator-norm sampling rule.

\prlsubheading{Average infidelity}
Let $\rho_\psi:=\ket{\psi}\!\bra{\psi}$ and let
$F(\rho,\sigma):=\norm{\sqrt{\rho}\sqrt{\sigma}}_1^2$ denote the state
fidelity.  We define the infidelity between the output states of the qDRIFT and
ideal channels, averaged over Haar-random input states, as
\begin{equation}
    \overline r_p(t,N)
    :=\E_{\psi\sim\mathrm{Haar}}
      \left[
      1-F\!\left(
        \mathcal{E}_{p,t}^{(N)}(\rho_\psi),
        \mathcal{U}_t(\rho_\psi)
      \right)
      \right].
    \label{eq:average-output-state-infidelity}
\end{equation}
Define the operator-valued variance
\begin{equation}
    D_p:=\sum_{j=1}^{K}\frac{H_j^2}{p_j}-H^2.
    \label{eq:variance-operator}
\end{equation}
The nonasymptotic estimate derived in the Appendix gives
\begin{equation}
    \overline r_p(t,N)
    \leq\frac{t^2}{Nd}\operatorname{Tr}(D_p),
    \label{eq:average-infidelity-bound}
\end{equation}
where $d=2^n$ is the Hilbert-space dimension.
The distribution dependence of the common error parameter is
\begin{align}
    \frac{1}{d}\operatorname{Tr}(D_p)
    &=\sum_{j=1}^{K}\frac{\norm{H_j}_{\mathrm{HS}}^2}{p_j}
      -\norm{H}_{\mathrm{HS}}^2.
    \label{eq:variance-trace}
\end{align}
Only the first term depends on $p$.  Applying the same Cauchy--Schwarz
argument as in Eq.~\eqref{eq:diamond-objective}, we obtain
\begin{equation}
    \sum_{j=1}^{K}\frac{\norm{H_j}_{\mathrm{HS}}^2}{p_j}
    \geq\left(\sum_{j=1}^{K}\norm{H_j}_{\mathrm{HS}}\right)^2,
\end{equation}
with equality for
\begin{equation}
    p_j^{(\mathrm{HS})}
    =\frac{\norm{H_j}_{\mathrm{HS}}}{\lambda_{\mathrm{HS}}},
    \qquad
    \lambda_{\mathrm{HS}}:=\sum_{k=1}^{K}\norm{H_k}_{\mathrm{HS}}.
    \label{eq:infidelity-optimal-distribution}
\end{equation}
Substituting Eq.~\eqref{eq:infidelity-optimal-distribution} gives
\begin{equation}
    \overline r_{p^{(\mathrm{HS})}}(t,N)
    \leq\frac{t^2}{N}
    \left[
      \lambda_{\mathrm{HS}}^2
      -\norm{H}_{\mathrm{HS}}^2
    \right].
    \label{eq:optimal-average-infidelity}
\end{equation}
\prlsubheading{Average time-evolution signal error}
For a pure input state $\ket{\psi}$, we define the complex time-evolution
signal
\begin{equation}
    g_\psi(t):=\bra{\psi}U_t\ket{\psi}.
    \label{eq:ideal-time-signal}
\end{equation}
The real and imaginary parts of this signal can be estimated using Hadamard
tests~\cite{Cleve1998}.  In single-ancilla phase estimation, signals measured
at a sequence of evolution times are processed classically to estimate
eigenenergies~\cite{Kitaev1995,Dobsicek2007,WiebeGranade2016,OBrien2019,LinTong2022,Gunther2026}.
Using the randomized Hamiltonian simulation above, each circuit repetition
applies the unitary
\begin{equation}
    V_{\boldsymbol{J},t}
    :=\prod_{k=1}^{N}
      \exp\!\left(-i\delta\frac{H_{J_k}}{p_{J_k}}\right),
    \label{eq:random-product-formula}
\end{equation}
where $\boldsymbol{J}=(J_1,\ldots,J_N)$ and the indices $J_k$ are sampled
independently from $p$.
Repeated Hadamard tests estimate
\begin{align}
    \widetilde g_{\psi,p}(t,N)
    &:=\E_{\boldsymbol{J}}
      \left[\bra{\psi}V_{\boldsymbol{J},t}\ket{\psi}\right]
      \nonumber\\
    &=\bra{\psi}\overline V_{p,t}^{(N)}\ket{\psi},
    \label{eq:randomized-time-signal}
\end{align}
where $\overline V_{p,t}^{(N)}:=\E_{\boldsymbol{J}}[V_{\boldsymbol{J},t}]$ is
an operator average, rather than a channel average.
We quantify the complex-plane bias of the signal for a fixed input by
$\abs{g_\psi(t)-\widetilde g_{\psi,p}(t,N)}$ and define the Haar-averaged signal
error as
\begin{equation}
    \overline\varepsilon_p(t,N)
    :=\E_{\psi\sim\mathrm{Haar}}
      \left[\abs{g_\psi(t)-\widetilde g_{\psi,p}(t,N)}\right].
    \label{eq:haar-averaged-signal-error}
\end{equation}
As shown in the Appendix, the same nonasymptotic estimate gives
\begin{equation}
    \overline\varepsilon_p(t,N)
    \leq\frac{t^2}{2Nd}\operatorname{Tr}(D_p).
    \label{eq:signal-bound}
\end{equation}
Consequently, the same distribution $p^{(\mathrm{HS})}$ minimizes the upper
bounds on both average infidelity and average signal error.
For the time signal, it gives
\begin{equation}
    \overline\varepsilon_{p^{(\mathrm{HS})}}(t,N)
    \leq\frac{t^2}{2N}
    \left[
      \lambda_{\mathrm{HS}}^2
      -\norm{H}_{\mathrm{HS}}^2
    \right].
    \label{eq:optimal-signal-error}
\end{equation}
Since $\operatorname{Tr}(D_p)$ is independent of $t$, the same distribution minimizes
the nonasymptotic upper bound at all query times.  Equations~\eqref{eq:average-infidelity-bound}
and~\eqref{eq:signal-bound} identify $\operatorname{Tr}(D_p)/d$ as a
common error parameter for the two tasks.

\begin{figure*}[t]
    \centering
    \begin{minipage}[t]{0.455\textwidth}
      \vspace{0pt}
      \centering
      \includegraphics[width=\linewidth]
        {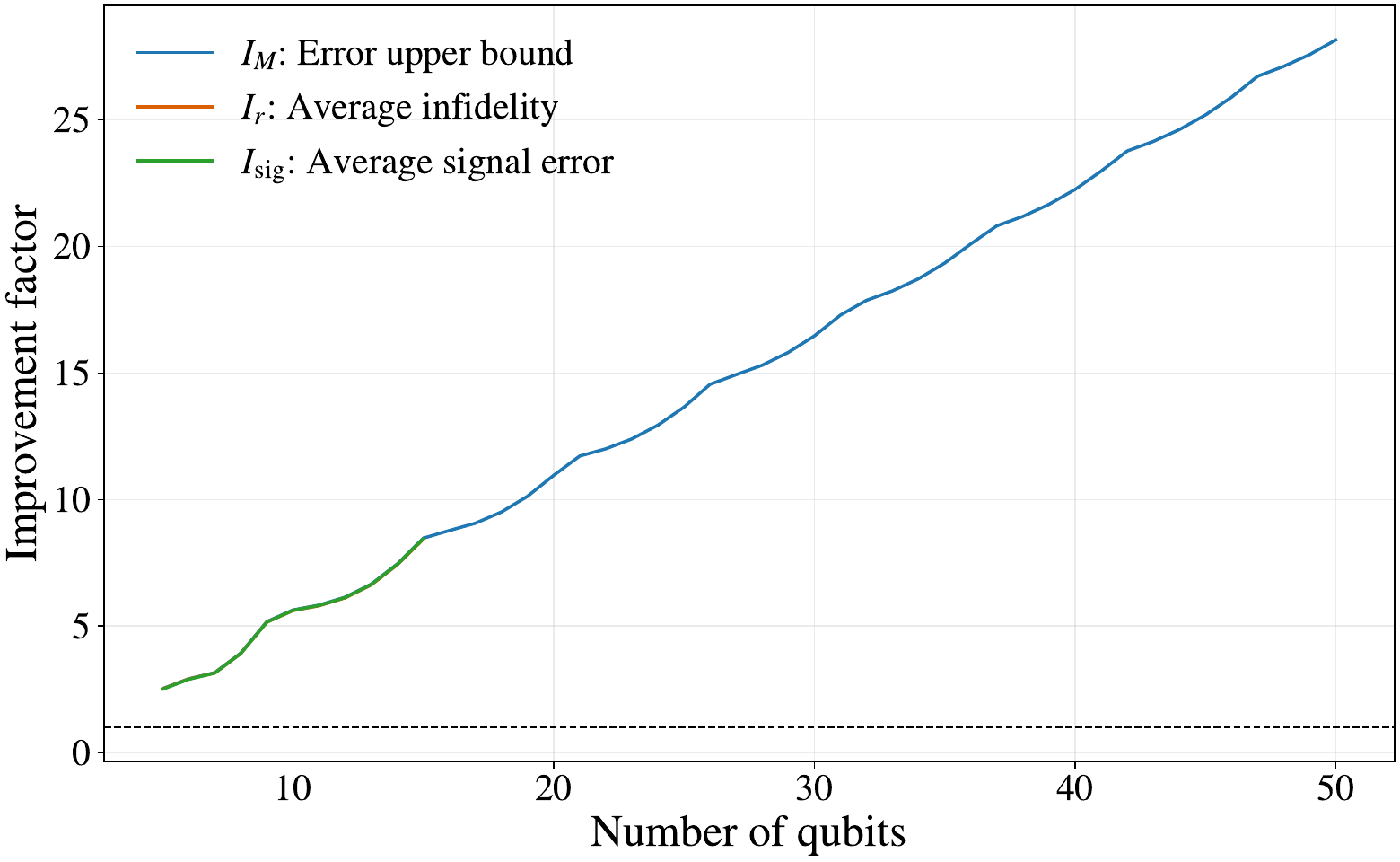}
      \textbf{(a)}
    \end{minipage}
    \hfill
    \begin{minipage}[t]{0.525\textwidth}
      \vspace{0pt}
      \centering
      \includegraphics[width=\linewidth]
        {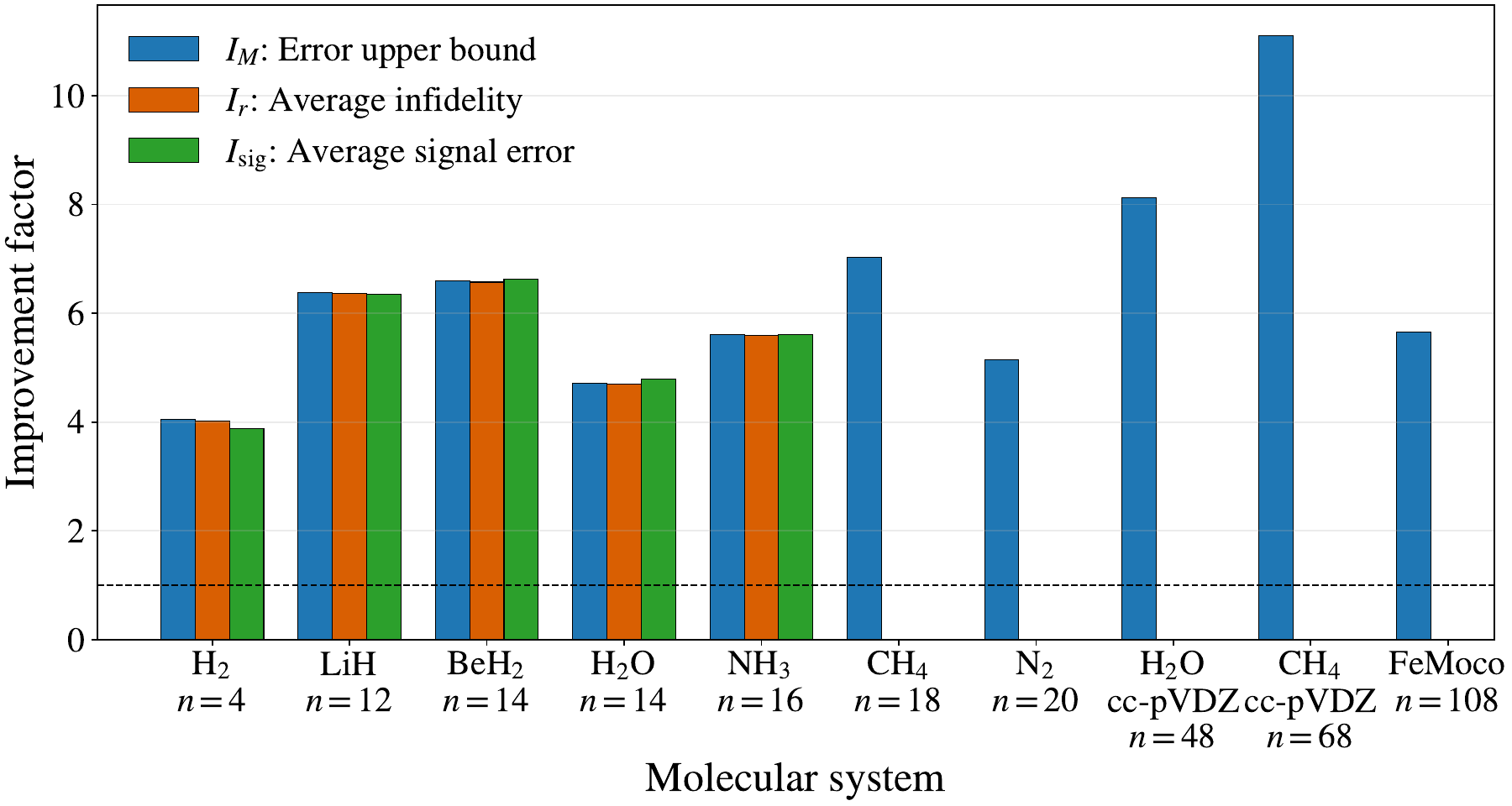}
      \textbf{(b)}
    \end{minipage}
    \caption{
      Improvement factors for (a) the quartic SYK model and (b) molecular
      electronic-structure Hamiltonians.  In (a), the solid curves show the
      medians over 50 independent coupling realizations with $J=1$ at each
      system size.
      The predicted factor $I_M$ is evaluated for $n=5,\ldots,50$, while the
      actual-error factors $I_r$ and $I_{\mathrm{sig}}$ are evaluated for the
      classically tractable sizes $n\leq15$.  In (b), $I_M$ is shown for
      every molecular instance, whereas $I_r$ and $I_{\mathrm{sig}}$ are
      evaluated through the 16-qubit $\mathrm{NH}_3$ instance.  For both
      panels, $t\lambda_{\mathrm{single}}=1$ and $N=100$.  The dashed
      lines denote no improvement.
    }
    \label{fig:improvement-factors}
\end{figure*}

\paragraph{Application to Pauli decompositions.---}

To make the distinction between the two sampling rules concrete, consider a
Hamiltonian expressed as a linear combination of $L$ distinct Pauli
strings,
\begin{equation}
    H=\sum_{\ell=1}^{L}c_\ell P_\ell.
    \label{eq:pauli-hamiltonian}
\end{equation}
If every Pauli term is sampled separately, $H_\ell=c_\ell P_\ell$, then
$\norm{H_\ell}_{\mathrm{op}}=\norm{H_\ell}_{\mathrm{HS}}=\abs{c_\ell}$.  Hence the conventional
and average-error-optimized distributions coincide:
\begin{equation}
    p_\ell^{(\mathrm{op})}=p_\ell^{(\mathrm{HS})}
    =\frac{\abs{c_\ell}}{\lambda_{\mathrm{single}}},
    \;
    \lambda_{\mathrm{single}}:=\norm{\boldsymbol{c}}_1=\sum_{\ell=1}^{L}\abs{c_\ell}.
    \label{eq:ungrouped-pauli-distribution}
\end{equation}
Thus changing the error criterion alone does not modify qDRIFT for a
term-by-term Pauli decomposition.  For this distribution,
Eq.~\eqref{eq:variance-trace} gives the common average-error parameter
\begin{equation}
    \frac{1}{d}\operatorname{Tr}(D_p)
    =\lambda_{\mathrm{single}}^2-\norm{\boldsymbol{c}}_2^2.
    \label{eq:ungrouped-pauli-average-error-parameter}
\end{equation}
Hence its scale is set by $\lambda_{\mathrm{single}}^2$ whenever no small number
of coefficients dominates, so that
$\norm{\boldsymbol{c}}_2^2\ll\lambda_{\mathrm{single}}^2$.

This coefficient-$\ell_1$ scale can be reduced by grouping.  Suppose that the
Pauli strings are partitioned into $K$ disjoint groups
$G_1,\ldots,G_K$, and each block
$H_j:=\sum_{\ell\in G_j}c_\ell P_\ell$ is treated as one simulatable term. 
For example, choosing mutually commuting Pauli strings within each group allows the block evolution to be implemented as a product of Pauli rotations.
By Pauli orthogonality, $\norm{H_j}_{\mathrm{HS}}=\norm{\boldsymbol{c}_{G_j}}_2$,
where $\norm{\boldsymbol{c}_{G_j}}_2:=\sqrt{\sum_{\ell\in G_j}c_\ell^2}$.
The average-error-optimized rule therefore becomes
\begin{equation}
    p_j^{(\mathrm{HS})}
    =\frac{\norm{\boldsymbol{c}_{G_j}}_2}
      {\lambda_{\mathrm{group}}},
    \quad
    \lambda_{\mathrm{group}}
    :=\sum_j\norm{\boldsymbol{c}_{G_j}}_2.
    \label{eq:grouped-pauli-hs-distribution}
\end{equation}
In this sense, grouping replaces the coefficient $\ell_1$ weight appearing in
the usual term-by-term qDRIFT construction by a blockwise coefficient
$\ell_2$ weight.  At the optimum, the common average-error parameter is
\begin{equation}
    \frac{1}{d}\operatorname{Tr}(D_{p^{(\mathrm{HS})}})
    =\lambda_{\mathrm{group}}^2-\norm{\boldsymbol{c}}_2^2.
    \label{eq:grouped-pauli-average-error-parameter}
\end{equation}
Indeed,
$\lambda_{\mathrm{group}}=\sum_j\norm{\boldsymbol{c}_{G_j}}_2
\leq\sum_j\norm{\boldsymbol{c}_{G_j}}_1
=\lambda_{\mathrm{single}}$.
For singleton groups, Eq.~\eqref{eq:grouped-pauli-average-error-parameter}
reduces to Eq.~\eqref{eq:ungrouped-pauli-average-error-parameter}, whereas
combining terms replaces $\ell_1$ norms by $\ell_2$ norms
within each block and can therefore lower the bound.  
Our bounds apply to any
grouping for which the block evolutions can be implemented.
Once the grouping is fixed, the sampling weights and the variance parameter
can be computed in $O(L)$ classical time.

\paragraph{Benchmarks for fermionic Hamiltonians.---}

Using our average-error analysis, we numerically
demonstrate reduced qDRIFT errors for fermionic Hamiltonians.  We consider quartic
Sachdev--Ye--Kitaev (SYK) models and molecular electronic-structure
Hamiltonians mapped to qubits by the Jordan--Wigner transformation.  We
compare two qDRIFT implementations: one samples individual Pauli terms,
and the other samples mutually commuting groups of Pauli terms.  For the
latter, we use a grouping inspired by Baranyai
decompositions~\cite{CsakanyThom2023}.  For dense quartic interactions, a
typical group contains $\Theta(n)$ Pauli terms.  The groups are chosen so
that each grouped evolution has $R_z$ depth one, as for a
single Pauli term.  Thus, at the same number of qDRIFT steps $N$, both
methods have the same $R_z$ depth.
Hamiltonian construction, grouping, and error-estimation details are given in
the Appendix.

The conventional method samples individual Pauli strings according to
Eq.~\eqref{eq:ungrouped-pauli-distribution}, whereas the grouped method uses
Eq.~\eqref{eq:grouped-pauli-hs-distribution}.  We quantify the improvement in
the error bounds by
\begin{equation}
    I_M
    :=\frac{\operatorname{Tr}(D_{\mathrm{single}})}
      {\operatorname{Tr}(D_{\mathrm{group}})}
    =\frac{\lambda_{\mathrm{single}}^2-\norm{\boldsymbol c}_2^2}
      {\lambda_{\mathrm{group}}^2-\norm{\boldsymbol c}_2^2},
    \label{eq:predicted-improvement-factor}
\end{equation}
where $I_M>1$ indicates an improvement.  Because the prefactors in
Eqs.~\eqref{eq:average-infidelity-bound} and~\eqref{eq:signal-bound}
cancel, $I_M$ is the same predicted improvement factor for both error
measures.  We also evaluate the actual-error improvement factors
\begin{equation}
    I_r:=\frac{\overline r_{\mathrm{single}}(t,N)}
      {\overline r_{\mathrm{group}}(t,N)},
    \qquad
    I_{\mathrm{sig}}
    :=\frac{\overline\varepsilon_{\mathrm{single}}(t,N)}
      {\overline\varepsilon_{\mathrm{group}}(t,N)}.
    \label{eq:actual-improvement-factors}
\end{equation}
The Haar averages entering $I_r$ and $I_{\mathrm{sig}}$ are estimated by
Monte Carlo sampling; estimator details and sample counts are given in the
Appendix.

Figure~\ref{fig:improvement-factors}(a) shows that $I_M$ grows approximately
linearly with $n$.  The SYK couplings have a common statistical scale.
Together with the group sizes described above, this gives a typical
within-group norm ratio
$\norm{\boldsymbol c_{G_j}}_1/\norm{\boldsymbol c_{G_j}}_2
=\Theta(\sqrt n)$.
The resulting reduction in the summed norms, combined with the quadratic
dependence of the common error parameter on this sum, explains the observed
approximately linear growth of $I_M$.
For the classically accessible sizes $n\leq15$, both $I_r$ and
$I_{\mathrm{sig}}$ nearly coincide with $I_M$.  Thus, in this regime,
$\operatorname{Tr}(D_p)/d$ accurately predicts the relative error reduction
produced by grouping for both error measures.

Figure~\ref{fig:improvement-factors}(b) shows that grouping also improves the
common error bound for every molecular Hamiltonian considered, with $I_M$
ranging from about $4$ to $11$.  The largest improvement occurs for
$\mathrm{CH}_4$ in the cc-pVDZ basis.  The variation is not monotonic in the
number of qubits---in particular, the FeMoco value is below those of both
cc-pVDZ instances---showing that the gain is controlled by the coefficient
distribution within the chosen groups rather than by system size alone.  For
the instances up to 16 qubits, the actual factors $I_r$ and
$I_{\mathrm{sig}}$ nearly coincide with each other and closely track $I_M$.
Thus, the common error parameter also captures the improvement from grouping
for both error measures in molecular Hamiltonians.

\paragraph{Discussion and conclusion.---}

We have derived nonasymptotic bounds on the Haar-averaged infidelity and
time-evolution signal error of importance-sampled qDRIFT, both governed by
the common variance parameter $\operatorname{Tr}(D_p)/d$.  For a fixed
Hamiltonian decomposition, both bounds are minimized by sampling in
proportion to the normalized Hilbert--Schmidt norms of the blocks.
The resulting average-error bounds are controlled by
$\lambda_{\mathrm{HS}}$ rather than the $\lambda_{\mathrm{op}}$ governing
the conventional worst-case bound.  For Hamiltonians expressed as linear
combinations of Pauli strings, grouping replaces the termwise coefficient-$\ell_1$ weight
by a sum of blockwise coefficient-$\ell_2$ weights.  Our SYK and molecular
benchmarks show reductions in the error bounds at fixed $R_z$ depth.
For small systems, we further confirm that the improvement factors in the
error bounds closely match those in the actual errors.
These results establish a connection
between the error criterion, the sampling probabilities, and the choice of
Hamiltonian blocks in randomized simulation.

Future work includes applying this framework to other Hamiltonian classes,
such as local spin systems and lattice fermion models, and optimizing the
grouping itself.  Since larger blocks can require more costly circuits,
a useful objective is to jointly optimize grouping and sampling for a
specified resource measure, accounting for Clifford gates and rotation
synthesis as well as $R_z$ depth.  Extending the analysis beyond Haar-random
inputs to low-energy or fixed-particle-number ensembles is another promising
direction.  Such extensions would connect the sampling strategy more
directly to the input states used in many-body dynamics and quantum chemistry.

\begin{acknowledgments}

\paragraph*{Acknowledgments.---}
This work is supported by MEXT Quantum Leap Flagship Program (MEXT Q-LEAP) Grant No. JPMXS0120319794, JST COI-NEXT Grant No. JPMJPF2014, JST Moonshot R\&D Grant No. JPMJMS256E, and JST CREST JPMJCR24I3.
H.M. is supported by JSPS KAKENHI Grant No. JP26KJ1663.

\end{acknowledgments}

\paragraph*{Data availability.---}
The code used for the numerical experiments in this work is
publicly available on GitHub~\cite{qDRIFTCode}.


\IfFileExists{references.bib}{\bibliography{references}}{}

\begin{thebibliography}{31}%
\makeatletter
\providecommand \@ifxundefined [1]{%
 \@ifx{#1\undefined}
}%
\providecommand \@ifnum [1]{%
 \ifnum #1\expandafter \@firstoftwo
 \else \expandafter \@secondoftwo
 \fi
}%
\providecommand \@ifx [1]{%
 \ifx #1\expandafter \@firstoftwo
 \else \expandafter \@secondoftwo
 \fi
}%
\providecommand \natexlab [1]{#1}%
\providecommand \enquote  [1]{``#1''}%
\providecommand \bibnamefont  [1]{#1}%
\providecommand \bibfnamefont [1]{#1}%
\providecommand \citenamefont [1]{#1}%
\providecommand \href@noop [0]{\@secondoftwo}%
\providecommand \href [0]{\begingroup \@sanitize@url \@href}%
\providecommand \@href[1]{\@@startlink{#1}\@@href}%
\providecommand \@@href[1]{\endgroup#1\@@endlink}%
\providecommand \@sanitize@url [0]{\catcode `\\12\catcode `\$12\catcode `\&12\catcode `\#12\catcode `\^12\catcode `\_12\catcode `\%12\relax}%
\providecommand \@@startlink[1]{}%
\providecommand \@@endlink[0]{}%
\providecommand \url  [0]{\begingroup\@sanitize@url \@url }%
\providecommand \@url [1]{\endgroup\@href {#1}{\urlprefix }}%
\providecommand \urlprefix  [0]{URL }%
\providecommand \Eprint [0]{\href }%
\providecommand \doibase [0]{https://doi.org/}%
\providecommand \selectlanguage [0]{\@gobble}%
\providecommand \bibinfo  [0]{\@secondoftwo}%
\providecommand \bibfield  [0]{\@secondoftwo}%
\providecommand \translation [1]{[#1]}%
\providecommand \BibitemOpen [0]{}%
\providecommand \bibitemStop [0]{}%
\providecommand \bibitemNoStop [0]{.\EOS\space}%
\providecommand \EOS [0]{\spacefactor3000\relax}%
\providecommand \BibitemShut  [1]{\csname bibitem#1\endcsname}%
\let\auto@bib@innerbib\@empty
\bibitem [{\citenamefont {Feynman}(1982)}]{Feynman1982}%
  \BibitemOpen
  \bibfield  {author} {\bibinfo {author} {\bibfnamefont {R.~P.}\ \bibnamefont {Feynman}},\ }\bibfield  {title} {\bibinfo {title} {Simulating physics with computers},\ }\href {https://doi.org/10.1007/BF02650179} {\bibfield  {journal} {\bibinfo  {journal} {International Journal of Theoretical Physics}\ }\textbf {\bibinfo {volume} {21}},\ \bibinfo {pages} {467} (\bibinfo {year} {1982})}\BibitemShut {NoStop}%
\bibitem [{\citenamefont {Lloyd}(1996)}]{Lloyd1996}%
  \BibitemOpen
  \bibfield  {author} {\bibinfo {author} {\bibfnamefont {S.}~\bibnamefont {Lloyd}},\ }\bibfield  {title} {\bibinfo {title} {Universal quantum simulators},\ }\href {https://doi.org/10.1126/science.273.5278.1073} {\bibfield  {journal} {\bibinfo  {journal} {Science}\ }\textbf {\bibinfo {volume} {273}},\ \bibinfo {pages} {1073} (\bibinfo {year} {1996})}\BibitemShut {NoStop}%
\bibitem [{\citenamefont {Reiher}\ \emph {et~al.}(2017)\citenamefont {Reiher}, \citenamefont {Wiebe}, \citenamefont {Svore}, \citenamefont {Wecker},\ and\ \citenamefont {Troyer}}]{Reiher2017}%
  \BibitemOpen
  \bibfield  {author} {\bibinfo {author} {\bibfnamefont {M.}~\bibnamefont {Reiher}}, \bibinfo {author} {\bibfnamefont {N.}~\bibnamefont {Wiebe}}, \bibinfo {author} {\bibfnamefont {K.~M.}\ \bibnamefont {Svore}}, \bibinfo {author} {\bibfnamefont {D.}~\bibnamefont {Wecker}},\ and\ \bibinfo {author} {\bibfnamefont {M.}~\bibnamefont {Troyer}},\ }\bibfield  {title} {\bibinfo {title} {Elucidating reaction mechanisms on quantum computers},\ }\href {https://doi.org/10.1073/pnas.1619152114} {\bibfield  {journal} {\bibinfo  {journal} {Proceedings of the National Academy of Sciences}\ }\textbf {\bibinfo {volume} {114}},\ \bibinfo {pages} {7555} (\bibinfo {year} {2017})}\BibitemShut {NoStop}%
\bibitem [{\citenamefont {Ge}\ \emph {et~al.}(2019)\citenamefont {Ge}, \citenamefont {Tura},\ and\ \citenamefont {Cirac}}]{Ge2019}%
  \BibitemOpen
  \bibfield  {author} {\bibinfo {author} {\bibfnamefont {Y.}~\bibnamefont {Ge}}, \bibinfo {author} {\bibfnamefont {J.}~\bibnamefont {Tura}},\ and\ \bibinfo {author} {\bibfnamefont {J.~I.}\ \bibnamefont {Cirac}},\ }\bibfield  {title} {\bibinfo {title} {Faster ground state preparation and high-precision ground energy estimation with fewer qubits},\ }\href {https://doi.org/10.1063/1.5027484} {\bibfield  {journal} {\bibinfo  {journal} {Journal of Mathematical Physics}\ }\textbf {\bibinfo {volume} {60}},\ \bibinfo {pages} {022202} (\bibinfo {year} {2019})}\BibitemShut {NoStop}%
\bibitem [{\citenamefont {Poulin}\ and\ \citenamefont {Wocjan}(2009)}]{PoulinWocjan2009}%
  \BibitemOpen
  \bibfield  {author} {\bibinfo {author} {\bibfnamefont {D.}~\bibnamefont {Poulin}}\ and\ \bibinfo {author} {\bibfnamefont {P.}~\bibnamefont {Wocjan}},\ }\bibfield  {title} {\bibinfo {title} {Sampling from the thermal quantum {G}ibbs state and evaluating partition functions with a quantum computer},\ }\href {https://doi.org/10.1103/PhysRevLett.103.220502} {\bibfield  {journal} {\bibinfo  {journal} {Physical Review Letters}\ }\textbf {\bibinfo {volume} {103}},\ \bibinfo {pages} {220502} (\bibinfo {year} {2009})}\BibitemShut {NoStop}%
\bibitem [{\citenamefont {Kitaev}(1995)}]{Kitaev1995}%
  \BibitemOpen
  \bibfield  {author} {\bibinfo {author} {\bibfnamefont {A.~Y.}\ \bibnamefont {Kitaev}},\ }\href {https://doi.org/10.48550/arXiv.quant-ph/9511026} {\bibinfo {title} {Quantum measurements and the {A}belian {S}tabilizer {P}roblem}} (\bibinfo {year} {1995}),\ \Eprint {https://arxiv.org/abs/quant-ph/9511026} {arXiv:quant-ph/9511026} \BibitemShut {NoStop}%
\bibitem [{\citenamefont {Cleve}\ \emph {et~al.}(1998)\citenamefont {Cleve}, \citenamefont {Ekert}, \citenamefont {Macchiavello},\ and\ \citenamefont {Mosca}}]{Cleve1998}%
  \BibitemOpen
  \bibfield  {author} {\bibinfo {author} {\bibfnamefont {R.}~\bibnamefont {Cleve}}, \bibinfo {author} {\bibfnamefont {A.}~\bibnamefont {Ekert}}, \bibinfo {author} {\bibfnamefont {C.}~\bibnamefont {Macchiavello}},\ and\ \bibinfo {author} {\bibfnamefont {M.}~\bibnamefont {Mosca}},\ }\bibfield  {title} {\bibinfo {title} {Quantum algorithms revisited},\ }\href {https://doi.org/10.1098/rspa.1998.0164} {\bibfield  {journal} {\bibinfo  {journal} {Proceedings of the Royal Society A: Mathematical, Physical and Engineering Sciences}\ }\textbf {\bibinfo {volume} {454}},\ \bibinfo {pages} {339} (\bibinfo {year} {1998})}\BibitemShut {NoStop}%
\bibitem [{\citenamefont {Abrams}\ and\ \citenamefont {Lloyd}(1999)}]{AbramsLloyd1999}%
  \BibitemOpen
  \bibfield  {author} {\bibinfo {author} {\bibfnamefont {D.~S.}\ \bibnamefont {Abrams}}\ and\ \bibinfo {author} {\bibfnamefont {S.}~\bibnamefont {Lloyd}},\ }\bibfield  {title} {\bibinfo {title} {Quantum algorithm providing exponential speed increase for finding eigenvalues and eigenvectors},\ }\href {https://doi.org/10.1103/PhysRevLett.83.5162} {\bibfield  {journal} {\bibinfo  {journal} {Physical Review Letters}\ }\textbf {\bibinfo {volume} {83}},\ \bibinfo {pages} {5162} (\bibinfo {year} {1999})}\BibitemShut {NoStop}%
\bibitem [{\citenamefont {Berry}\ \emph {et~al.}(2015)\citenamefont {Berry}, \citenamefont {Childs}, \citenamefont {Cleve}, \citenamefont {Kothari},\ and\ \citenamefont {Somma}}]{Berry2015}%
  \BibitemOpen
  \bibfield  {author} {\bibinfo {author} {\bibfnamefont {D.~W.}\ \bibnamefont {Berry}}, \bibinfo {author} {\bibfnamefont {A.~M.}\ \bibnamefont {Childs}}, \bibinfo {author} {\bibfnamefont {R.}~\bibnamefont {Cleve}}, \bibinfo {author} {\bibfnamefont {R.}~\bibnamefont {Kothari}},\ and\ \bibinfo {author} {\bibfnamefont {R.~D.}\ \bibnamefont {Somma}},\ }\bibfield  {title} {\bibinfo {title} {Simulating {H}amiltonian dynamics with a truncated {T}aylor series},\ }\href {https://doi.org/10.1103/PhysRevLett.114.090502} {\bibfield  {journal} {\bibinfo  {journal} {Physical Review Letters}\ }\textbf {\bibinfo {volume} {114}},\ \bibinfo {pages} {090502} (\bibinfo {year} {2015})}\BibitemShut {NoStop}%
\bibitem [{\citenamefont {Low}\ and\ \citenamefont {Chuang}(2019)}]{LowChuang2019}%
  \BibitemOpen
  \bibfield  {author} {\bibinfo {author} {\bibfnamefont {G.~H.}\ \bibnamefont {Low}}\ and\ \bibinfo {author} {\bibfnamefont {I.~L.}\ \bibnamefont {Chuang}},\ }\bibfield  {title} {\bibinfo {title} {Hamiltonian simulation by qubitization},\ }\href {https://doi.org/10.22331/q-2019-07-12-163} {\bibfield  {journal} {\bibinfo  {journal} {Quantum}\ }\textbf {\bibinfo {volume} {3}},\ \bibinfo {pages} {163} (\bibinfo {year} {2019})}\BibitemShut {NoStop}%
\bibitem [{\citenamefont {Campbell}(2019)}]{Campbell2019}%
  \BibitemOpen
  \bibfield  {author} {\bibinfo {author} {\bibfnamefont {E.}~\bibnamefont {Campbell}},\ }\bibfield  {title} {\bibinfo {title} {Random {C}ompiler for {F}ast {H}amiltonian {S}imulation},\ }\href {https://doi.org/10.1103/PhysRevLett.123.070503} {\bibfield  {journal} {\bibinfo  {journal} {Physical Review Letters}\ }\textbf {\bibinfo {volume} {123}},\ \bibinfo {pages} {070503} (\bibinfo {year} {2019})}\BibitemShut {NoStop}%
\bibitem [{\citenamefont {Low}\ and\ \citenamefont {Chuang}(2017)}]{LowChuang2017}%
  \BibitemOpen
  \bibfield  {author} {\bibinfo {author} {\bibfnamefont {G.~H.}\ \bibnamefont {Low}}\ and\ \bibinfo {author} {\bibfnamefont {I.~L.}\ \bibnamefont {Chuang}},\ }\bibfield  {title} {\bibinfo {title} {Optimal {H}amiltonian simulation by quantum signal processing},\ }\href {https://doi.org/10.1103/PhysRevLett.118.010501} {\bibfield  {journal} {\bibinfo  {journal} {Physical Review Letters}\ }\textbf {\bibinfo {volume} {118}},\ \bibinfo {pages} {010501} (\bibinfo {year} {2017})}\BibitemShut {NoStop}%
\bibitem [{\citenamefont {Low}\ \emph {et~al.}(2019)\citenamefont {Low}, \citenamefont {Kliuchnikov},\ and\ \citenamefont {Wiebe}}]{LowKliuchnikovWiebe2019}%
  \BibitemOpen
  \bibfield  {author} {\bibinfo {author} {\bibfnamefont {G.~H.}\ \bibnamefont {Low}}, \bibinfo {author} {\bibfnamefont {V.}~\bibnamefont {Kliuchnikov}},\ and\ \bibinfo {author} {\bibfnamefont {N.}~\bibnamefont {Wiebe}},\ }\href {https://doi.org/10.48550/arXiv.1907.11679} {\bibinfo {title} {Well-conditioned multiproduct {H}amiltonian simulation}} (\bibinfo {year} {2019}),\ \Eprint {https://arxiv.org/abs/1907.11679} {arXiv:1907.11679 [quant-ph]} \BibitemShut {NoStop}%
\bibitem [{\citenamefont {Childs}\ \emph {et~al.}(2019)\citenamefont {Childs}, \citenamefont {Ostrander},\ and\ \citenamefont {Su}}]{ChildsOstranderSu2019}%
  \BibitemOpen
  \bibfield  {author} {\bibinfo {author} {\bibfnamefont {A.~M.}\ \bibnamefont {Childs}}, \bibinfo {author} {\bibfnamefont {A.}~\bibnamefont {Ostrander}},\ and\ \bibinfo {author} {\bibfnamefont {Y.}~\bibnamefont {Su}},\ }\bibfield  {title} {\bibinfo {title} {Faster quantum simulation by randomization},\ }\href {https://doi.org/10.22331/q-2019-09-02-182} {\bibfield  {journal} {\bibinfo  {journal} {Quantum}\ }\textbf {\bibinfo {volume} {3}},\ \bibinfo {pages} {182} (\bibinfo {year} {2019})}\BibitemShut {NoStop}%
\bibitem [{\citenamefont {Hagan}\ and\ \citenamefont {Wiebe}(2023)}]{HaganWiebe2023}%
  \BibitemOpen
  \bibfield  {author} {\bibinfo {author} {\bibfnamefont {M.}~\bibnamefont {Hagan}}\ and\ \bibinfo {author} {\bibfnamefont {N.}~\bibnamefont {Wiebe}},\ }\bibfield  {title} {\bibinfo {title} {Composite quantum simulations},\ }\href {https://doi.org/10.22331/q-2023-11-14-1181} {\bibfield  {journal} {\bibinfo  {journal} {Quantum}\ }\textbf {\bibinfo {volume} {7}},\ \bibinfo {pages} {1181} (\bibinfo {year} {2023})}\BibitemShut {NoStop}%
\bibitem [{\citenamefont {Nakaji}\ \emph {et~al.}(2024)\citenamefont {Nakaji}, \citenamefont {Bagherimehrab},\ and\ \citenamefont {Aspuru-Guzik}}]{Nakaji2024}%
  \BibitemOpen
  \bibfield  {author} {\bibinfo {author} {\bibfnamefont {K.}~\bibnamefont {Nakaji}}, \bibinfo {author} {\bibfnamefont {M.}~\bibnamefont {Bagherimehrab}},\ and\ \bibinfo {author} {\bibfnamefont {A.}~\bibnamefont {Aspuru-Guzik}},\ }\bibfield  {title} {\bibinfo {title} {High-order randomized compiler for {H}amiltonian simulation},\ }\href {https://doi.org/10.1103/PRXQuantum.5.020330} {\bibfield  {journal} {\bibinfo  {journal} {PRX Quantum}\ }\textbf {\bibinfo {volume} {5}},\ \bibinfo {pages} {020330} (\bibinfo {year} {2024})}\BibitemShut {NoStop}%
\bibitem [{\citenamefont {Kiss}\ \emph {et~al.}(2023)\citenamefont {Kiss}, \citenamefont {Grossi},\ and\ \citenamefont {Roggero}}]{Kiss2023}%
  \BibitemOpen
  \bibfield  {author} {\bibinfo {author} {\bibfnamefont {O.}~\bibnamefont {Kiss}}, \bibinfo {author} {\bibfnamefont {M.}~\bibnamefont {Grossi}},\ and\ \bibinfo {author} {\bibfnamefont {A.}~\bibnamefont {Roggero}},\ }\bibfield  {title} {\bibinfo {title} {Importance sampling for stochastic quantum simulations},\ }\href {https://doi.org/10.22331/q-2023-04-13-977} {\bibfield  {journal} {\bibinfo  {journal} {Quantum}\ }\textbf {\bibinfo {volume} {7}},\ \bibinfo {pages} {977} (\bibinfo {year} {2023})}\BibitemShut {NoStop}%
\bibitem [{\citenamefont {Wu}\ \emph {et~al.}(2026)\citenamefont {Wu}, \citenamefont {Fan},\ and\ \citenamefont {Zhang}}]{Wu2026}%
  \BibitemOpen
  \bibfield  {author} {\bibinfo {author} {\bibfnamefont {Y.-X.}\ \bibnamefont {Wu}}, \bibinfo {author} {\bibfnamefont {Y.-Z.}\ \bibnamefont {Fan}},\ and\ \bibinfo {author} {\bibfnamefont {D.-B.}\ \bibnamefont {Zhang}},\ }\bibfield  {title} {\bibinfo {title} {Fluctuation-guided adaptive random compiler for {H}amiltonian simulation},\ }\href {https://doi.org/10.1103/52wr-1hys} {\bibfield  {journal} {\bibinfo  {journal} {Physical Review Applied}\ }\textbf {\bibinfo {volume} {25}},\ \bibinfo {pages} {054034} (\bibinfo {year} {2026})}\BibitemShut {NoStop}%
\bibitem [{\citenamefont {Schubert}\ and\ \citenamefont {Mendl}(2023)}]{SchubertMendl2023}%
  \BibitemOpen
  \bibfield  {author} {\bibinfo {author} {\bibfnamefont {A.}~\bibnamefont {Schubert}}\ and\ \bibinfo {author} {\bibfnamefont {C.~B.}\ \bibnamefont {Mendl}},\ }\bibfield  {title} {\bibinfo {title} {Trotter error with commutator scaling for the {F}ermi--{H}ubbard model},\ }\href {https://doi.org/10.1103/PhysRevB.108.195105} {\bibfield  {journal} {\bibinfo  {journal} {Physical Review B}\ }\textbf {\bibinfo {volume} {108}},\ \bibinfo {pages} {195105} (\bibinfo {year} {2023})}\BibitemShut {NoStop}%
\bibitem [{\citenamefont {Zhao}\ \emph {et~al.}(2022)\citenamefont {Zhao}, \citenamefont {Zhou}, \citenamefont {Shaw}, \citenamefont {Li},\ and\ \citenamefont {Childs}}]{Zhao2022}%
  \BibitemOpen
  \bibfield  {author} {\bibinfo {author} {\bibfnamefont {Q.}~\bibnamefont {Zhao}}, \bibinfo {author} {\bibfnamefont {Y.}~\bibnamefont {Zhou}}, \bibinfo {author} {\bibfnamefont {A.~F.}\ \bibnamefont {Shaw}}, \bibinfo {author} {\bibfnamefont {T.}~\bibnamefont {Li}},\ and\ \bibinfo {author} {\bibfnamefont {A.~M.}\ \bibnamefont {Childs}},\ }\bibfield  {title} {\bibinfo {title} {Hamiltonian simulation with random inputs},\ }\href {https://doi.org/10.1103/PhysRevLett.129.270502} {\bibfield  {journal} {\bibinfo  {journal} {Physical Review Letters}\ }\textbf {\bibinfo {volume} {129}},\ \bibinfo {pages} {270502} (\bibinfo {year} {2022})}\BibitemShut {NoStop}%
\bibitem [{\citenamefont {Chen}\ and\ \citenamefont {Brand{\~a}o}(2024)}]{ChenBrandao2024}%
  \BibitemOpen
  \bibfield  {author} {\bibinfo {author} {\bibfnamefont {C.-F.}\ \bibnamefont {Chen}}\ and\ \bibinfo {author} {\bibfnamefont {F.~G. S.~L.}\ \bibnamefont {Brand{\~a}o}},\ }\bibfield  {title} {\bibinfo {title} {Average-case speedup for product formulas},\ }\href {https://doi.org/10.1007/s00220-023-04912-5} {\bibfield  {journal} {\bibinfo  {journal} {Communications in Mathematical Physics}\ }\textbf {\bibinfo {volume} {405}},\ \bibinfo {pages} {32} (\bibinfo {year} {2024})}\BibitemShut {NoStop}%
\bibitem [{\citenamefont {Dob{\v{s}}{\'i}{\v{c}}ek}\ \emph {et~al.}(2007)\citenamefont {Dob{\v{s}}{\'i}{\v{c}}ek}, \citenamefont {Johansson}, \citenamefont {Shumeiko},\ and\ \citenamefont {Wendin}}]{Dobsicek2007}%
  \BibitemOpen
  \bibfield  {author} {\bibinfo {author} {\bibfnamefont {M.}~\bibnamefont {Dob{\v{s}}{\'i}{\v{c}}ek}}, \bibinfo {author} {\bibfnamefont {G.}~\bibnamefont {Johansson}}, \bibinfo {author} {\bibfnamefont {V.}~\bibnamefont {Shumeiko}},\ and\ \bibinfo {author} {\bibfnamefont {G.}~\bibnamefont {Wendin}},\ }\bibfield  {title} {\bibinfo {title} {Arbitrary accuracy iterative quantum phase estimation algorithm using a single ancillary qubit: A two-qubit benchmark},\ }\href {https://doi.org/10.1103/PhysRevA.76.030306} {\bibfield  {journal} {\bibinfo  {journal} {Physical Review A}\ }\textbf {\bibinfo {volume} {76}},\ \bibinfo {pages} {030306} (\bibinfo {year} {2007})}\BibitemShut {NoStop}%
\bibitem [{\citenamefont {Wiebe}\ and\ \citenamefont {Granade}(2016)}]{WiebeGranade2016}%
  \BibitemOpen
  \bibfield  {author} {\bibinfo {author} {\bibfnamefont {N.}~\bibnamefont {Wiebe}}\ and\ \bibinfo {author} {\bibfnamefont {C.}~\bibnamefont {Granade}},\ }\bibfield  {title} {\bibinfo {title} {Efficient {B}ayesian phase estimation},\ }\href {https://doi.org/10.1103/PhysRevLett.117.010503} {\bibfield  {journal} {\bibinfo  {journal} {Physical Review Letters}\ }\textbf {\bibinfo {volume} {117}},\ \bibinfo {pages} {010503} (\bibinfo {year} {2016})}\BibitemShut {NoStop}%
\bibitem [{\citenamefont {O'Brien}\ \emph {et~al.}(2019)\citenamefont {O'Brien}, \citenamefont {Tarasinski},\ and\ \citenamefont {Terhal}}]{OBrien2019}%
  \BibitemOpen
  \bibfield  {author} {\bibinfo {author} {\bibfnamefont {T.~E.}\ \bibnamefont {O'Brien}}, \bibinfo {author} {\bibfnamefont {B.}~\bibnamefont {Tarasinski}},\ and\ \bibinfo {author} {\bibfnamefont {B.~M.}\ \bibnamefont {Terhal}},\ }\bibfield  {title} {\bibinfo {title} {Quantum phase estimation of multiple eigenvalues for small-scale (noisy) experiments},\ }\href {https://doi.org/10.1088/1367-2630/aafb8e} {\bibfield  {journal} {\bibinfo  {journal} {New Journal of Physics}\ }\textbf {\bibinfo {volume} {21}},\ \bibinfo {pages} {023022} (\bibinfo {year} {2019})}\BibitemShut {NoStop}%
\bibitem [{\citenamefont {Lin}\ and\ \citenamefont {Tong}(2022)}]{LinTong2022}%
  \BibitemOpen
  \bibfield  {author} {\bibinfo {author} {\bibfnamefont {L.}~\bibnamefont {Lin}}\ and\ \bibinfo {author} {\bibfnamefont {Y.}~\bibnamefont {Tong}},\ }\bibfield  {title} {\bibinfo {title} {Heisenberg-limited ground-state energy estimation for early fault-tolerant quantum computers},\ }\href {https://doi.org/10.1103/PRXQuantum.3.010318} {\bibfield  {journal} {\bibinfo  {journal} {PRX Quantum}\ }\textbf {\bibinfo {volume} {3}},\ \bibinfo {pages} {010318} (\bibinfo {year} {2022})}\BibitemShut {NoStop}%
\bibitem [{\citenamefont {G{\"u}nther}\ \emph {et~al.}(2026)\citenamefont {G{\"u}nther}, \citenamefont {Witteveen}, \citenamefont {Schmidhuber}, \citenamefont {Miller}, \citenamefont {Christandl},\ and\ \citenamefont {Harrow}}]{Gunther2026}%
  \BibitemOpen
  \bibfield  {author} {\bibinfo {author} {\bibfnamefont {J.}~\bibnamefont {G{\"u}nther}}, \bibinfo {author} {\bibfnamefont {F.}~\bibnamefont {Witteveen}}, \bibinfo {author} {\bibfnamefont {A.}~\bibnamefont {Schmidhuber}}, \bibinfo {author} {\bibfnamefont {M.}~\bibnamefont {Miller}}, \bibinfo {author} {\bibfnamefont {M.}~\bibnamefont {Christandl}},\ and\ \bibinfo {author} {\bibfnamefont {A.~W.}\ \bibnamefont {Harrow}},\ }\bibfield  {title} {\bibinfo {title} {Phase estimation with partially randomized time evolution},\ }\href {https://doi.org/10.1103/ynxb-p2xq} {\bibfield  {journal} {\bibinfo  {journal} {PRX Quantum}\ }\textbf {\bibinfo {volume} {7}},\ \bibinfo {pages} {020332} (\bibinfo {year} {2026})}\BibitemShut {NoStop}%
\bibitem [{\citenamefont {Sachdev}\ and\ \citenamefont {Ye}(1993)}]{SachdevYe1993}%
  \BibitemOpen
  \bibfield  {author} {\bibinfo {author} {\bibfnamefont {S.}~\bibnamefont {Sachdev}}\ and\ \bibinfo {author} {\bibfnamefont {J.}~\bibnamefont {Ye}},\ }\bibfield  {title} {\bibinfo {title} {Gapless spin-fluid ground state in a random quantum {H}eisenberg magnet},\ }\href {https://doi.org/10.1103/PhysRevLett.70.3339} {\bibfield  {journal} {\bibinfo  {journal} {Physical Review Letters}\ }\textbf {\bibinfo {volume} {70}},\ \bibinfo {pages} {3339} (\bibinfo {year} {1993})}\BibitemShut {NoStop}%
\bibitem [{\citenamefont {Maldacena}\ and\ \citenamefont {Stanford}(2016)}]{MaldacenaStanford2016}%
  \BibitemOpen
  \bibfield  {author} {\bibinfo {author} {\bibfnamefont {J.}~\bibnamefont {Maldacena}}\ and\ \bibinfo {author} {\bibfnamefont {D.}~\bibnamefont {Stanford}},\ }\bibfield  {title} {\bibinfo {title} {Remarks on the {Sachdev--Ye--Kitaev} model},\ }\href {https://doi.org/10.1103/PhysRevD.94.106002} {\bibfield  {journal} {\bibinfo  {journal} {Physical Review D}\ }\textbf {\bibinfo {volume} {94}},\ \bibinfo {pages} {106002} (\bibinfo {year} {2016})}\BibitemShut {NoStop}%
\bibitem [{\citenamefont {Csakany}\ and\ \citenamefont {Thom}(2023)}]{CsakanyThom2023}%
  \BibitemOpen
  \bibfield  {author} {\bibinfo {author} {\bibfnamefont {B.}~\bibnamefont {Csakany}}\ and\ \bibinfo {author} {\bibfnamefont {A.~J.~W.}\ \bibnamefont {Thom}},\ }\href {https://doi.org/10.48550/arXiv.2311.11826} {\bibinfo {title} {Optimised {B}aranyai partitioning of the second quantised {H}amiltonian}} (\bibinfo {year} {2023}),\ \Eprint {https://arxiv.org/abs/2311.11826} {arXiv:2311.11826 [quant-ph]} \BibitemShut {NoStop}%
\bibitem [{qDR()}]{qDRIFTCode}%
  \BibitemOpen
  \href {https://github.com/sakimori-coder/Optimized-Randomized-Hamiltonian-Simulation-via-Average-Error-Analysis} {\bibinfo {title} {{Optimized Randomized Hamiltonian Simulation via Average-Error Analysis}}},\ \bibinfo {howpublished} {GitHub repository}\BibitemShut {NoStop}%
\bibitem [{\citenamefont {Sun}\ \emph {et~al.}(2020)\citenamefont {Sun}, \citenamefont {Zhang}, \citenamefont {Banerjee} \emph {et~al.}}]{Sun2020PySCF}%
  \BibitemOpen
  \bibfield  {author} {\bibinfo {author} {\bibfnamefont {Q.}~\bibnamefont {Sun}}, \bibinfo {author} {\bibfnamefont {X.}~\bibnamefont {Zhang}}, \bibinfo {author} {\bibfnamefont {S.}~\bibnamefont {Banerjee}}, \emph {et~al.},\ }\bibfield  {title} {\bibinfo {title} {Recent developments in the {PySCF} program package},\ }\href {https://doi.org/10.1063/5.0006074} {\bibfield  {journal} {\bibinfo  {journal} {The Journal of Chemical Physics}\ }\textbf {\bibinfo {volume} {153}},\ \bibinfo {pages} {024109} (\bibinfo {year} {2020})}\BibitemShut {NoStop}%
\end{thebibliography}%


\appendix
\allowdisplaybreaks[1]

\section{Derivation of the nonasymptotic average-error bounds}
\label{app:average-error-bounds}

We prove Eqs.~\eqref{eq:average-infidelity-bound} and~\eqref{eq:signal-bound}
using a common bound on the mean time-evolution operator.
Throughout, $\norm{A}_1:=\operatorname{Tr}\sqrt{A^\dagger A}$ denotes the
trace norm, and $\E_\psi$ denotes the average over Haar-random input states.

\prlsubheading{Mean time-evolution operator}
The common estimate is
\begin{equation}
    \norm{\overline V_{p,t}^{(N)}-U_t}_1
    \leq\frac{t^2}{2N}\operatorname{Tr}(D_p),
    \label{eq:appendix-global-trace-bound}
\end{equation}
which holds for every real $t$ and positive integer $N$.
To prove it, let $\delta=t/N$ and set
\begin{equation}
    X_j:=\frac{H_j}{p_j},
    \qquad
    A_j:=X_j-H,
    \qquad
    V_{j,s}:=e^{-isX_j}.
\end{equation}
Since $\sum_jp_jX_j=H$ and $\sum_jp_j=1$, we have
$\sum_jp_jA_j=0$ and
\begin{align}
    \sum_jp_jA_j^2
    &=\sum_jp_jX_j^2-\sum_jp_j(X_jH+HX_j)+H^2
      \nonumber\\
    &=\sum_j\frac{H_j^2}{p_j}-H^2
      =D_p\succeq0.
    \label{eq:appendix-centered-variance}
\end{align}
No commutativity assumption is needed for this identity.
We first take $\delta\geq0$.  The derivative identity
\begin{equation*}
    \frac{\mathrm{d}}{\mathrm{d}s}(U_{\delta-s}V_{j,s})
    =-iU_{\delta-s}A_jV_{j,s}
\end{equation*}
gives Duhamel's formula upon integration:
\begin{equation}
    V_{j,\delta}-U_\delta
    =-i\int_0^\delta U_{\delta-s}A_jV_{j,s}\,\mathrm{d}s.
    \label{eq:appendix-duhamel}
\end{equation}
Substituting the same identity for $V_{j,s}-U_s$ into
Eq.~\eqref{eq:appendix-duhamel} yields
\begin{align}
    V_{j,\delta}-U_\delta
    &=-i\int_0^\delta U_{\delta-s}A_jU_s\,\mathrm{d}s
      \nonumber\\
    &\quad-\int_0^\delta\!\mathrm{d}s\int_0^s\!\mathrm{d}r\,
      U_{\delta-s}A_jU_{s-r}A_jV_{j,r}.
    \label{eq:appendix-double-duhamel}
\end{align}
Define $\overline V_{p,\delta}:=\sum_jp_jV_{j,\delta}$ as the mean one-step
time-evolution operator.  Averaging Eq.~\eqref{eq:appendix-double-duhamel}
over $j$ cancels the first integral exactly, since $\sum_jp_jA_j=0$:
\begin{align}
    \overline V_{p,\delta}-U_\delta
    &=-\sum_{j=1}^{K}p_j
      \int_0^\delta\!\mathrm{d}s\int_0^s\!\mathrm{d}r\,
      \nonumber\\
    &\qquad\times U_{\delta-s}A_jU_{s-r}A_jV_{j,r}.
    \label{eq:appendix-one-step-difference}
\end{align}
Unitary invariance and H\"older's inequality for Schatten norms imply
\begin{align}
    \norm{U_{\delta-s}A_jU_{s-r}A_jV_{j,r}}_1
    &=\norm{A_jU_{s-r}A_j}_1
      \nonumber\\
    &\leq d\norm{A_j}_{\mathrm{HS}}\norm{U_{s-r}A_j}_{\mathrm{HS}}
      \nonumber\\
    &=d\norm{A_j}_{\mathrm{HS}}^2
      =\operatorname{Tr}(A_j^2).
\end{align}
The triangle inequality and $D_p=\sum_jp_jA_j^2$ then give
\begin{align}
    \norm{\overline V_{p,\delta}-U_\delta}_1
    &\leq\sum_jp_j\int_0^\delta\!\mathrm{d}s
      \int_0^s\!\mathrm{d}r\,\operatorname{Tr}(A_j^2)
      \nonumber\\
    &=\frac{\delta^2}{2}\sum_jp_j\operatorname{Tr}(A_j^2)
      \nonumber\\
    &=\frac{\delta^2}{2}\operatorname{Tr}(D_p).
    \label{eq:appendix-one-step-trace-bound}
\end{align}
Taking the average before the norm is what retains the centered variance
$D_p$, including the subtraction of $H^2$.
The same bound holds for negative $\delta$, since
$\overline V_{p,-\delta}=\overline V_{p,\delta}^\dagger$, $U_{-\delta}=U_\delta^\dagger$, and the
trace norm is invariant under taking the adjoint.

As an average of unitary operators, $\overline V_{p,\delta}$ is a contraction,
$\norm{\overline V_{p,\delta}}_{\mathrm{op}}\leq1$.
Independence of the sampled steps gives
$\overline V_{p,t}^{(N)}=\bigl(\overline V_{p,\delta}\bigr)^N$, while $U_t=U_\delta^N$.
The telescoping identity, which does not require commutativity, is
\begin{equation}
    \overline V_{p,\delta}^N-U_\delta^N
    =\sum_{k=0}^{N-1}\overline V_{p,\delta}^{N-1-k}
      (\overline V_{p,\delta}-U_\delta)U_\delta^k.
\end{equation}
Using $\norm{BAC}_1\leq\norm{B}_{\mathrm{op}}\norm{A}_1
\norm{C}_{\mathrm{op}}$ on each term and $\delta=t/N$, we obtain
\begin{align*}
    \norm{\overline V_{p,t}^{(N)}-U_t}_1
    &\leq N\norm{\overline V_{p,\delta}-U_\delta}_1
      \\
    &\leq\frac{N\delta^2}{2}\operatorname{Tr}(D_p)
      =\frac{t^2}{2N}\operatorname{Tr}(D_p).
\end{align*}
This proves Eq.~\eqref{eq:appendix-global-trace-bound} without a
short-time approximation or an asymptotic remainder.

\prlsubheading{Average infidelity}
For each sampled sequence, write
$a_{\psi,\boldsymbol{J}}:=\bra{\psi}U_t^\dagger
V_{\boldsymbol{J},t}\ket{\psi}$.
Because the ideal output is pure and
$\mathcal{E}_{p,t}^{(N)}(\rho_\psi)
=\E_{\boldsymbol{J}}[V_{\boldsymbol{J},t}\rho_\psi
V_{\boldsymbol{J},t}^\dagger]$, the fidelity is linear in the randomized
output state.  Hence
\begin{equation}
    \overline r_p(t,N)
    =\E_{\psi,\boldsymbol{J}}
      \left[1-\abs{a_{\psi,\boldsymbol{J}}}^2\right].
\end{equation}
For any complex number $a$,
$1-\abs{a}^2=2(1-\Re a)-\abs{1-a}^2\leq2(1-\Re a)$.
Using the Haar first moment $\E_\psi[\rho_\psi]=I/d$ and
$\E_{\boldsymbol{J}}[V_{\boldsymbol{J},t}]
=\overline V_{p,t}^{(N)}$, we therefore find
\begin{align}
    \overline r_p(t,N)
    &\leq2\left(1-\Re\E_{\psi,\boldsymbol{J}}
      [a_{\psi,\boldsymbol{J}}]\right)
      \nonumber\\
    &=\frac{2}{d}\Re\operatorname{Tr}\!\left[
      U_t^\dagger(U_t-\overline V_{p,t}^{(N)})\right]
      \nonumber\\
    &\leq\frac{2}{d}\norm{\overline V_{p,t}^{(N)}-U_t}_1
      \nonumber\\
    &\leq\frac{t^2}{Nd}\operatorname{Tr}(D_p).
    \label{eq:appendix-average-infidelity-bound}
\end{align}
The second inequality uses $\abs{\operatorname{Tr}(A)}\leq\norm{A}_1$
and unitary invariance of the trace norm; the last inequality is
Eq.~\eqref{eq:appendix-global-trace-bound}.
This proves Eq.~\eqref{eq:average-infidelity-bound}.

\prlsubheading{Average time-evolution signal error}
We first show that, for any operator $A$,
\begin{equation}
    \E_\psi\abs{\bra{\psi}A\ket{\psi}}
    \leq\frac{\norm{A}_1}{d}.
    \label{eq:appendix-haar-trace-norm}
\end{equation}
Let $A=\sum_a s_a\ket{u_a}\!\bra{v_a}$ be a singular-value decomposition,
where $s_a\geq0$ and the left and right singular vectors are normalized.
No Hermiticity or normality assumption on $A$ is needed.
The triangle and Cauchy--Schwarz inequalities give
\begin{align}
    \E_\psi\abs{\bra{\psi}A\ket{\psi}}
    &\leq\sum_a s_a\E_\psi
      \left[\abs{\braket{\psi|u_a}}\abs{\braket{v_a|\psi}}\right]
      \nonumber\\
    &\leq\sum_a s_a
      \sqrt{\E_\psi\abs{\braket{\psi|u_a}}^2
        \E_\psi\abs{\braket{v_a|\psi}}^2}
      \nonumber\\
    &=\frac{1}{d}\sum_a s_a=\frac{\norm{A}_1}{d},
\end{align}
since $\E_\psi[\rho_\psi]=I/d$ makes each squared-overlap average equal to
$1/d$.  Taking $A=U_t-\overline V_{p,t}^{(N)}$ yields
\begin{align}
    \overline\varepsilon_p(t,N)
    &=\E_\psi\abs{\bra{\psi}
      (U_t-\overline V_{p,t}^{(N)})\ket{\psi}}
      \nonumber\\
    &\leq\frac{1}{d}\norm{U_t-\overline V_{p,t}^{(N)}}_1
      \nonumber\\
    &\leq\frac{t^2}{2Nd}\operatorname{Tr}(D_p),
    \label{eq:appendix-average-signal-bound}
\end{align}
where the last step uses Eq.~\eqref{eq:appendix-global-trace-bound}.
This proves Eq.~\eqref{eq:signal-bound}.  Both bounds hold for every positive
integer $N$, with no asymptotic remainder.
The circuit average is taken before the absolute value, so this bound controls
the bias of the mean signal, not the average error of individual trajectories
or the statistical error from finitely many measurements.
The two main-text bounds use only $\E_\psi[\rho_\psi]=I/d$ and therefore also
hold for any pure-state ensemble with this first moment.

\section{Commuting Pauli groups with $R_z$ depth one}
\label{app:grouping}

We describe the grouping used in the benchmarks.  It uses disjoint-orbital
packing inspired by Baranyai grouping~\cite{CsakanyThom2023} to construct
blocks whose time evolutions have $R_z$ depth one.
The numerical implementation uses a greedy packing, rather than an exact
Baranyai factorization.

\prlsubheading{Depth-one condition}
For each block $H_j=\sum_{\ell\in G_j}c_\ell P_\ell$, we require its
time evolution to be implementable by a Clifford transformation, one
layer of parallel single-qubit $Z$ rotations, and the inverse Clifford
transformation.  Formally, for every real $\theta$, there exist a Clifford
unitary $C_j$, real
parameters $\theta_\ell$, and pairwise distinct qubit indices
$q_\ell\in\{1,\ldots,n\}$ for $\ell\in G_j$ such that
\begin{equation}
    e^{-i\theta H_j}
    =C_j^\dagger\left[
      \prod_{\ell\in G_j}
      e^{-i\theta_\ell Z_{q_\ell}}
      \right]C_j.
    \label{eq:appendix-group-depth-one}
\end{equation}
This implementation is obtained from a Clifford transformation satisfying
\begin{equation*}
    C_jP_\ell C_j^\dagger=s_\ell Z_{q_\ell},
    \qquad \ell\in G_j,
\end{equation*}
where $s_\ell=\pm1$.  Such a transformation exists if and only if the Pauli
strings in $G_j$ are mutually commuting and independent modulo phase.
Here, independence
means that no nonempty product of distinct strings in the group equals
$\pm I$.
Commutativity alone does not ensure this condition: for example,
$Z_1,Z_2,Z_1Z_2$ commute, but their product is $I$.
The resource comparison counts $R_z$ layers, not the Clifford gates or
their routing overhead.

\prlsubheading{Four-orbital packets and packing}
For four distinct Jordan--Wigner sites $S=\{p,q,r,s\}$ with $p<q<r<s$,
the canonical quartic strings have $X$ or $Y$ on $S$ and the common parity
string $\Pi_S=\prod_{p<a<q}Z_a\prod_{r<a<s}Z_a$.
For even $Y$ parity, we split the eight possible endpoint words into
\begin{align*}
    \mathcal A_{\mathrm e}&=\{XXXX,YYXX,YXYX,YXXY\},\\
    \mathcal B_{\mathrm e}&=\{YYYY,XXYY,XYXY,XYYX\}.
\end{align*}
For odd $Y$ parity, which also occurs in the SYK Hamiltonians, the two
packets are
\begin{align*}
    \mathcal A_{\mathrm o}&=\{YXXX,XYXX,XXYX,XXXY\},\\
    \mathcal B_{\mathrm o}&=\{XYYY,YXYY,YYXY,YYYX\}.
\end{align*}
Each word is placed on $(p,q,r,s)$ and multiplied by $\Pi_S$; only
strings present in the Hamiltonian are retained.  Each packet is
commuting and independent, whereas the full eight-string set at fixed
parity is dependent and is generated, modulo phase, by four independent
strings.  Packets on disjoint endpoint sets can be combined
while preserving both properties, even when their Jordan--Wigner parity
strings overlap.

Packets are processed in decreasing order of their coefficient $\ell_1$
weights.  Each is added to an existing group with a disjoint endpoint set,
favoring larger groups, or starts a new group if none of the inspected
candidates is compatible.  The implementation inspects up to 16 candidate
groups of each eligible size.
Remaining strings, including the diagonal and lower-orbital sectors, are
first collected by their set of $X/Y$ sites and the parity of their number
of $Y$ factors.  These sets commute.  Within each set, strings are processed
lexicographically and assigned to a group only if the addition preserves
independence; up to eight open, nonfull groups are tested
before opening another.  No Pauli terms are discarded by grouping.

For a dense four-orbital sector with $4\mid n$, a complete Baranyai
construction partitions the orbital quartets into $\Theta(n^3)$ matchings
containing $n/4$ quartets each.  Other sizes can be padded with at most
three dummy sites and the dummy-containing quartets subsequently removed.
Splitting each quartet into the constant number of packets
above preserves this scaling and gives groups of $\Theta(n)$ Pauli terms.
This motivates the greedy construction used here; an exact factorization
or an optimal group count is not assumed for the numerical packing.

\section{Numerical methods and parameters}
\label{app:numerics}

\prlsubheading{Molecular Hamiltonians}
The molecular instances are specified in
Table~\ref{tab:appendix-molecular-settings}.  Except for FeMoco, they are
neutral singlets, with Hartree--Fock molecular orbitals and molecular
integrals obtained using PySCF~\cite{Sun2020PySCF}.
For these instances, the full orbital space is retained, without frozen orbitals.
For FeMoco we use the supplied 54-electron, 54-spatial-orbital active-space
FCIDUMP integrals associated with Ref.~\cite{Reiher2017}, giving 108 qubits.
No new geometry optimization or mean-field calculation is performed for
FeMoco.  All molecular Hamiltonians are converted to Pauli strings using
our implementation of the Jordan--Wigner transformation.
After combining identical Pauli strings, molecular coefficients with
$|c_\ell|\leq10^{-12}$ Hartree are omitted.  The identity component is
removed before grouping, computing coefficient norms, and simulating
dynamics.  Its known phase could be restored to both the ideal and
simulated time signals without changing their absolute difference.

\begin{table}[tb]
    \caption{Molecular benchmark settings.  Bond lengths are in
    \AA{} and bond angles in degrees.  The two basis choices for water
    and methane use the same geometry.  FeMoco uses the supplied
    CAS(54e,54o) integrals rather than a new molecular calculation.}
    \label{tab:appendix-molecular-settings}
    \begin{ruledtabular}
    \begin{tabular}{lcl}
        Molecule & $n$ & Geometry \\
        \hline
        $\mathrm H_2$ & 4 & H--H: 0.735 \\
        LiH & 12 & Li--H: 1.45 \\
        $\mathrm{BeH}_2$ & 14 & Linear; Be--H: 1.3264 \\
        $\mathrm H_2\mathrm O$ & 14, 48 & O--H: 0.9576; H--O--H: 104.5 \\
        $\mathrm{NH}_3$ & 16 & N--H: 1.012; H--N--H: 106.7 \\
        $\mathrm{CH}_4$ & 18, 68 & Tetrahedral; C--H: 1.087 \\
        $\mathrm N_2$ & 20 & N--N: 1.0977 \\
        FeMoco & 108 & Supplied active-space integrals \\
    \end{tabular}
    \end{ruledtabular}
    \smallskip
    All non-FeMoco instances use STO-3G, except the 48-qubit water and
    68-qubit methane instances, which use cc-pVDZ.
\end{table}

\prlsubheading{Error estimation}
To estimate $I_r$ and $I_{\mathrm{sig}}$, we sample $S$ independent
Haar-random input states $\ket{\psi_a}$ in the full $2^n$-dimensional
Hilbert space and $R$ independent qDRIFT trajectories per input for each
decomposition.  Let
$\ket{\phi_a}=U_t\ket{\psi_a}$ and
$\ket{\chi_{ab}}=V_{\boldsymbol J_{ab},t}\ket{\psi_a}$
denote the ideal and trajectory outputs, respectively.  The error
estimators are
\begin{align}
    \widehat r
    &=\frac{1}{S}\sum_{a=1}^{S}
      \left(1-\frac{1}{R}\sum_{b=1}^{R}
      \abs{\braket{\phi_a|\chi_{ab}}}^2\right),
      \label{eq:appendix-infidelity-estimator}\\
    \widehat\varepsilon
    &=\frac{1}{S}\sum_{a=1}^{S}
      \left|\braket{\psi_a|\phi_a}
      -\frac{1}{R}\sum_{b=1}^{R}
       \braket{\psi_a|\chi_{ab}}\right|.
      \label{eq:appendix-signal-estimator}
\end{align}
The same input states and ideal outputs are reused for the two
decompositions, with independent trajectories for each method.  The
estimates above are substituted into
Eq.~\eqref{eq:actual-improvement-factors} to obtain $I_r$ and
$I_{\mathrm{sig}}$.
For the data in Fig.~\ref{fig:improvement-factors}, we use $S=100$ and
$R=1000$ for both Hamiltonian families.

\end{document}